\documentstyle[12pt,epsfig]{article}
\def\lsim{\raise0.3ex\hbox{$<$\kern-0.75em\raise-1.1ex\hbox{$\sim$}}}
\def\gsim{\raise0.3ex\hbox{$>$\kern-0.75em\raise-1.1ex\hbox{$\sim$}}}

\def\beq{\begin{equation}}
\def\eeq{\end{equation}}
\def\bea{\begin{eqnarray}}
\def\eea{\end{eqnarray}}
\def\bq{\begin{quote}}
\def\eq{\end{quote}}

\newcommand{\rr}{\mbox{\boldmath $r$}}

\def\gappeq{\mathrel{\rlap {\raise.5ex\hbox{$>$}}
{\lower.5ex\hbox{$\sim$}}}}

\def\lappeq{\mathrel{\rlap{\raise.5ex\hbox{$<$}}
{\lower.5ex\hbox{$\sim$}}}}

\def\Toprel#1\over#2{\mathrel{\mathop{#2}\limits^{#1}}}

\begin{document}
\pagestyle{empty}
\begin{center}
{\bf A PHENOMENOLOGICAL ANALYSIS OF THE LONGITUDINAL STRUCTURE FUNCTION AT SMALL  $x$ AND LOW $Q^2$}
\\

\vspace*{1cm}
 V.P. Gon\c{c}alves $^{1}$, M.V.T. Machado  $^{2,\,3}$\\
\vspace{0.3cm}
{$^{1}$ Instituto de F\'{\i}sica e Matem\'atica,  Universidade
Federal de Pelotas\\
Caixa Postal 354, CEP 96010-090, Pelotas, RS, Brazil\\
$^{2}$ \rm Universidade Estadual do Rio Grande do Sul, UERGS \\
Engenharia de Bioprocessos e Biotecnologia, Bento Gon\c{c}alves, RS, Brazil\\
        $^{3}$ \rm High Energy Physics Phenomenology Group, GFPAE,  IF-UFRGS \\
Caixa Postal 15051, CEP 91501-970, Porto Alegre, RS, Brazil}\\
\vspace*{1cm}
{\bf ABSTRACT}
\end{center}

\vspace*{1cm} \noindent

\vspace*{1.5cm} 

\vspace{-2.5cm} \setcounter{page}{1} \pagestyle{plain}

The longitudinal structure function in deep inelastic
scattering is one of  the observables from which the gluon
distribution can be unfolded. Consequently, this observable can be
used to constrain the QCD dynamics at small $x$. In this work we
compare the predictions of distinct QCD models with the recent
experimental results for $F_L(x,Q^2)$ at small $x$ and low $Q^2$
obtained by the H1 collaboration.  
We focus mainly on the color dipole approach, 
selecting those models which include saturation effects. Such  models are  suitable at this kinematical region and also resum a wide class of higher twist contributions to the observables. Therefore, we investigate the influence of these corrections to $F_L$ in the present region of interest.

\vspace{1.5cm}

\section{Introduction}

The small $x$ regime  in deep inelastic scattering (DIS) is one of
the frontiers of Quantum Chromodynamics (QCD). This represents the
challenge of studying the interface between perturbative and
nonperturbative QCD, with the characteristic  feature that the 
transition is taken in a kinematical region where the strong
coupling constant $\alpha_s$ is small. That region has been
explored by the electron-proton ($ep$) scattering at HERA, which has shown a striking rise of the proton structure function $F_2(x,Q^2)$ for
values $x<10^{-2}$. This behavior implies that the cross
section increases faster that logarithmically with the energy,
violating the Froissart bound. Therefore, new dynamical non-linear QCD effects associated to the unitarity corrections are expected to slow down its further growth \cite{glr,muqiu,ayala1,mcl,bal,kov,iancu}.  The search of signatures for these effects has been an
active subject of research in the last years \cite{ayala2,ayala3,golecwus,bgbk,kowtea,pes,scaling,Caldwell_Mara,Kwien_Motyka,Mariotto_Machado,Favart_Machado,dipolos,prl,iancu_munier,Forshaw1,Forshaw2,jamal}.

In particular, it has  been observed that the HERA data at small $x$ and low $Q^2$ can be
successfully described with the help of saturation models \cite{golecwus,bgbk,kowtea,iancu_munier}. Moreover, the experimental results for the total cross section \cite{scaling} and also for the  inclusive charm production \cite{prl} present the property of  geometric scaling, which is one
of the main characteristics of the high density QCD approaches (For a recent review see e. g.  Ref. \cite{jamal_iancu}).
The saturation (non-linear QCD) approaches are characterized by a typical  scale, denoted the saturation scale $Q^2_{\mathrm{s}}(x)$, which is energy
dependent, and marks the transition between the  linear (leading twist) perturbative QCD regime and saturation domain. As current phenomenological saturation models have indicated that for the HERA domain the saturation scale is smaller than 2 GeV$^2$, we expect that the
signatures of the saturation effects becomes more evident in the
region of small $x$ and very low $Q^2$.
Furthermore, some of  these approaches contain
information of all orders in $1/Q^2$, namely they resum higher twist
contributions \cite{peters,gotsman}. These corrections should be important at the low $Q^2$ region, where the leading twist (DGLAP) approaches would be in the limit of their aplicability. Therefore, saturation models are quite suitable for the present phenomenological study of the longitudinal structure function at low $Q^2$.

In this work we analyze the behavior of the longitudinal
structure function $F_L(x,Q^2)$ in this kinematical regime. One  considers  several QCD theoretical approaches, focusing mainly on the saturation models. The predictions are compared with the recent (preliminary)
$F_L$ experimental results, as determined from the 1999 minimum bias and the 2000 shifted vertex H1 data \cite{h1_prel}. A comment  related to these results is in order here. The  
experimental determination of $F_L$ is difficult since it usually requires cross 
sections measurements at different values of  center of mass energy, 
implying a change of beam energies. An alternative possibility 
is to apply the radiation of a hard photon by the incoming electron. Such hard 
radiation results into an effective reduction of the center of mass energy. Several 
studies on the use of  such events to measure $F_L$ have  been carried out 
\cite{favart}. 
With  these measurements, which in principle could be performed in the near future, it may be possible to explore the structure of $F_L(x,Q^2)$ in the low $x$ range. However, currently, to  obtain the $F_L$ data, the H1 Collaboration has parametrized the  structure function $F_2$ taken 
only  data for $y < 0.35$, where the  contribution of $F_L$ is small.  
This parameterization was evolved in $Q^2$ according to the DGLAP evolution 
equations, providing  predictions for  $F_2$ in the 
high $y$ region which allowed, by subtraction of the contribution of $F_2$ 
to the cross section, the determination of the longitudinal structure function (For more detailed  discussions see Refs. \cite{thornefl,h1_prel}).
Therefore, the $F_L$ data only are obtained after the use of a procedure in the measurements of the total cross section.   

The paper is organized as follows.
 In the next section, we briefly review the theoretical description of $F_L$ in the  linear DGLAP (leading twist) approximation and  summarize the main expressions considering the color dipole approach. For the latter, we introduce two representative saturation models which have their phenomenological parameters well constrained from the current small $x$ experimental data. The comparison of the numerical results, further discussions and conclusions are presented in the last section.

\section{Theoretical description of $F_L$ at small $x$}

The longitudinal structure function
$F_L$  corresponds to the interaction of the longitudinally
polarized virtual photon in the one-photon-exchange mechanism of
lepton-nucleon scattering. It is a very interesting dynamical
quantity since, at least at low $x$, its dominant contribution comes from gluons. While in the naive parton model this structure
function vanishes, at leading-order in $\alpha_s(Q^2)$ it acquires a leading twist contribution. At small
$x$ this contribution is driven by the gluon through the $g \rightarrow q \overline{q}$ transition and, in fact, $F_L$ can be used  as a
very useful quantity for a direct measurement of the gluon
distribution in a nucleon. One can write the longitudinal
structure function $F_L$ in terms of the cross section for the
absorption of longitudinally polarized photons as
\begin{eqnarray}
F_L(x,Q^2)  =  \frac{Q^2(1-x)}{4\pi^2 \alpha_{em}}\,\sigma^{\gamma^* p}_L(x,Q^2)
\approx \frac{Q^2}{4\pi^2 \alpha_{em}}\,\sigma^{\gamma^* p}_L(x,Q^2)
\label{fl}
\end{eqnarray}
at small $x$. Therefore, accurate  measurements of
$F_L$ at low $x$ and/or $Q^2$ would be helpful to constrain the
physics in that kinematical region. In particular, we expect that
this observable may discriminate between the leading twist
predictions, which consider the collinear factorization and parton
distributions determined from global fits, and the predictions
from the saturation models which resum a class of higher twist
contributions at small $x$.

While the longitudinal structure function is  (at least
theoretically) fairly well understood at high $Q^2$ very little
(if anything) is known about its possible extrapolation towards
the region of low $Q^2$ and small $x$ \cite{Badelek} (For recent discussions see e. g. \cite{Kotikov,Kotikov2}). Theoretically, we have that in the limit
$Q^2 \rightarrow 0$ the structure function  $F_L$ has to vanish as
$Q^4$.  It reflects the simple physical fact that the total cross
section  $\sigma_L \approx F_L/Q^2$ describing the interaction  of
longitudinally polarized virtual photons  has to vanish  in the
real photoproduction limit. On the other hand, the leading twist DGLAP  MRST \cite{mrst} and CTEQ \cite{cteq} global fits require the gluon distribution  to be valencelike or negative at
small $x$ and low $Q^2$ in order to describe the experimental data, leading to $F_L$ being negative at the smallest $x-Q^2$. At that region, a comparison of the predictions  at LO, NLO and NNLO using
MRST partons shown a poor description of the experimental results \cite{thorne}. However, the description is improved if a $\ln\,(1/x)$ resummation is considered \cite{thorne2}. Here, it is important to emphasize that higher-twist
contributions are not considered in these analyzes. Nevertheless, in
the global fit of the existing light-targets DIS data at LO, NLO
and NNLO QCD approximations, Alekhin \cite{alekhin} has estimated the high-twist
contributions to the structure functions. It was verified that these
terms do not vanish up to NNLO (See Fig. 12 in that reference) and
gives important contributions at both small and large $x$ regions.
Although the expectation that higher twist  plays an important role
at very large $x$ is not new, the contribution of these terms in
the small $x$ region has been a subject of  discussion only in the
last years. For instance, in Ref. \cite{martin_plb} a simple
parameterization of the higher twist contribution to the $F_2$
structure function have been used, and it was found that for $x<0.5$
the resulting correction is small and negative but beyond 0.6
large and positive. In particular, the higher twist contributions
for $x<0.01$ are very small at HERA low $x$ domain. In other
words, the experimental results for the $F_2$ structure function
in principle can be described by a leading twist approximation.
However, this feature can also be explained as being due to the
almost complete cancellation of the twist-4 corrections to the
transverse and longitudinal structure functions \cite{peters}.
Therefore, only a direct analyzes of $F_L$ could  discriminate
between leading twist and higher-twist resummations.

In order to address these issues, in what follows we present representative theoretical approaches taking into account the usual DGLAP leading twist approximation and the  twist resummation rendered by the saturation models. At leading order, twist-two and in the infinite momentum frame
the longitudinal structure function can be expressed in terms of
the  Altarelli-Martinelli equation \cite{Altarelli}
\begin{eqnarray}
F_L(x,Q^2) = \frac{\alpha_s(Q^2)}{2\pi}\,x^2\, \int_x^1 \frac{dy}{y^3}\,\left[\frac{8}{3}\,F_2(y,Q^2) + 4\,\sum_q e_q^2 \,\left(1-\frac{x}{y}\right)\,y\,g(y,Q^2)\right]\,\,,
\label{flalta}
\end{eqnarray}
which  shows the dependence of $F_L$ on the strong constant
coupling and on the gluon density. At small $x$, the second term is the dominant one since it is driven by the gluon distribution. Consequently, Eq. (\ref{flalta}) can be reasonably approximated  by
$F_L \approx 0.3\,\frac{4 \alpha_s}{3 \pi}\, x\,g\,(2.5\,x,Q^2)$
\cite{Cooper_Sarkar}. This relation demonstrates  the close relation
between the longitudinal structure function and the gluon
distribution. In our further numerical calculations using the Altarelli-Martinelli equation, one considers as input  the  MRST2001(LO) \cite{mrst} and GRV98LO \cite{grv98} parton distributions. We use the GRV98 parameterization in order to compare the collinear approach with the experimental results for $Q^2 \le 1$ GeV$^2$.

In the proton rest frame, the DIS process can be seen as 
a succession in time of three factorisable subprocesses: i) 
the photon fluctuates in a quark-antiquark pair with transverse separation $r_{\perp}\sim 1/Q$ long after the interaction, ii) this 
color dipole interacts with the proton target, iii) the quark pair
annihilates in a virtual photon. The interaction $\gamma^*p$ is further
factorized in the simple formulation \cite{dipole},
\begin{eqnarray}
\sigma_{L,T}^{\gamma^*p}(x,Q^2)=\!\int dz \,d^2r_{\perp}
|\Psi_{L,T}(z,r_{\perp},Q^2)|^2
\,\sigma_{dip}(x,r_{\perp}),\nonumber
\end{eqnarray}
where $z$ is the longitudinal momentum fraction of the quark,
$x\simeq Q^2/ W_{\gamma p}^2$ is equivalent to the Bjorken
variable.  The photon wavefunctions $\Psi_{L,T}$ are determined
from light cone perturbation theory and read as
\begin{eqnarray}
 |\Psi_{T}|^2 & = &\!  \frac{6\alpha_{\mathrm{em}}}{4\,\pi^2} \,
 \sum_f e_f^2 \, \left\{[z^2 + (1-z)^2]\, \varepsilon^2 \,K_1^2(\varepsilon \,r_{\perp})
 +\,  m_f^2 \, \,K_0^2(\varepsilon\,r_{\perp})
 \right\}\label{wtrans} \nonumber \\
 |\Psi_{L}|^2 & = &\! \frac{6\alpha_{\mathrm{em}}}{\pi^2} \,
\sum_f e_f^2 \, \left\{Q^2 \,z^2 (1-z)^2
\,K_0^2(\varepsilon\,r_{\perp}) \right\}, \label{wlongs}
 \end{eqnarray}
where the auxiliary variable $\varepsilon^2=z(1-z)\,Q^2 + m^2_f$
depends on the quark mass, $m_f$. The $K_{0,1}$ are the McDonald
functions and the summation is performed over the quark flavors.

The dipole hadron cross section $\sigma_{dip}$  contains all
information about the target and the strong interaction physics.
There are several phenomenological implementations for this
quantity \cite{ayala3,golecwus,bgbk,kowtea,dipolos,iancu_munier}. The main feature of these approaches is
to be able to match the soft (low $Q^2$) and hard (large $Q^2$)
regimes in an unified way. In the present work, we follow the
quite successful saturation models \cite{golecwus,iancu_munier},
which interpolates between the small and large dipole
configurations, providing color transparency behavior,
$\sigma_{dip}\sim \rr^2$, as $r \gg Q_s$  and constant behavior at
large dipole separations $r < Q_s$. It is important to emphasize
that in  the dipole models at small $x$, both $F_L$ and $F_2$ are
governed by $\sigma_{dip}$ and therefore behave similarly.
In particular, $F_L$ should go to zero when $Q^2 \rightarrow 0$ at
low $x$ in the dipole picture since $|\Psi_L|^2 \propto Q^2$. The
parameters of the saturation models have been obtained from  phenomenological adjustments to small $x$ HERA data. As a first model, we present the analytically simple GBW model, which resembles the main features of the Glauber-Mueller resummation. Its phenomenological aplication has been successful in a wide class of processes with a photon probe (DIS, diffractive DIS, Deeply Virtual Compton Scattering, heavy-quark production, two-photon physics)\cite{golecwus,kowtea,pes,scaling,Caldwell_Mara,Kwien_Motyka,Mariotto_Machado,Favart_Machado,prl}. The parameterization for the
dipole cross section in this model takes the eikonal-like form,
\begin{eqnarray}
\sigma_{dip} (\tilde{x}, \,\rr^2)  =  \sigma_0 \, \left[\, 1- \exp
\left(-\frac{\,Q_s^2(x)\,\rr^2}{4} \right) \, \right]\,,
\hspace{1cm} Q_s^2(x)  =  \left( \frac{x_0}{\tilde{x}}
\right)^{\lambda} \,\,\mathrm{GeV}^2\,. \label{gbwdip}
\end{eqnarray}
 where the parameters were obtained from a fit to the HERA data producing $\sigma_0=23.03 \,(29.12)$ mb, $\lambda= 0.288 \, (0.277)$ and $x_0=3.04 \cdot 10^{-4} \, (0.41 \cdot 10^{-4})$ for a 3-flavor
(4-flavor) analysis~\cite{golecwus}. An
additional parameter is the effective light quark mass, $m_f=0.14$
GeV, which plays the role of a regulator for the photoproduction
($Q^2=0$) cross section.

An  important aspect of the saturation models is that they 
resum a class of higher twist contributions which should be
non-negligible in the low $Q^2$ regime \cite{peters,gotsman}. Consequently, in this
kinematical region we may expect a discrimination between the
twist two calculations, usually considered in the global fits of
the experimental data, and the saturation models. Some hints of the
differences between these models has been presented in Refs.
\cite{peters,gotsman}. In particular, the
twist expansion of the GBW model has been calculated, with the
different twist terms in the massless limit given by:
\begin{eqnarray}
\label{twist4l}
\hbox{Twist-4} \hspace{1cm} \sigma_L^{(4)} & = & \sigma_0 \sum_f e_f^2 \frac{\alpha_{em}}{\pi}
\left(- \frac{94}{75} \xi^2 + \frac{4}{5} \psi (3) \xi^2 -\frac{4}{5}
\xi^2 \ln(1/\xi) \right)\\
\label{twist6l}
\hbox{Twist-6} \hspace{1cm} \sigma_L^{(6)}  & = & \sigma_0 \sum_f e_f^2 \frac{\alpha_{em}}{\pi}
\left( \frac{654}{1225} \xi^3 - \frac{36}{35} \psi (4) \xi^3 +\frac{36}{35}
\xi^3 \ln(1/\xi) \right),
\\
\label{twist8l}
\hbox{Twist-8}\hspace{1cm}  \sigma_L^{(8)}  & = & \sigma_0 \sum_f e_f^2 \frac{\alpha_{em}}{\pi}
\left(- \frac{1636}{18375} \xi^4 + \frac{48}{175} \psi (5) \xi^4 -\frac{48}
{175}\xi^4 \ln(1/\xi) \right),
\end{eqnarray}
where $\xi = \frac{Q_s^2}{Q^2}$ is the scaling variable which appears in the geometric scaling property of the inclusive cross section and $\psi (x)$  is the digamma function. The results above can be contrasted to the leading twist result $\sigma_L^{(2)}\simeq \sigma_0 \sum_f e_f^2 \frac{\alpha_{em}}{\pi}\,\xi$.

The longitudinal twist-4 and twist-8 terms give sizeable negative corrections to the leading twist contribution, mainly at $\xi=\frac{Q_s^2}{Q^2}\approx 1$. It is expected that a precise low $Q^2$ measurements of $F_L$ at small-$x$ could reveal this important feature. This is investigated in the analysis presented here, considering the recent H1 preliminary data on the longitudinal structure function. It should be noticed although $F_2$ had been measured in this region with accurate precision, its longitudinal and transverse twist-4 contributions  have opposite signs and almost the same order of magnitude. Hence, they approximately cancel each other and produce  either a small twist-4 correction. Therefore, for the inclusive $F_2$ structure function the higher twist corrections are  hidden in the mismatch between the longitudinal and transverse higher twist corrections.

Despite the saturation model to be very successful in describing HERA data, its functional form is only an approximation of the theoretical non-linear QCD approaches. On the other hand, an analytical expression for the dipole cross section can be obtained within the BFKL formalism. Currently, intense theoretical studies has been performed towards an understanding of the BFKL approach in the border of the saturation region \cite{IANCUGEO,MUNIERWALLON}. In particular, the dipole cross section has been calculated in both LO  and NLO BFKL  approach in the geometric scaling region \cite{BFKLSCAL}. It reads as,
\begin{eqnarray}
\sigma_{dip}(x,\rr)=\sigma_0\,\left[\rr^2 Q_{\mathrm{sat}}^2(x)\right]^{\gamma_{\mathrm{sat}}}\,\exp\left[ -\frac{\ln^2\,\left(\rr^2 Q_{\mathrm{sat}}^2\right)}{2\,\beta \,\bar{\alpha}_sY}\right]\,,
\label{sigmabfkl}
\end{eqnarray}
where $\sigma_0  =2\pi R_p^2$ ($R_p$ is the proton radius) is the overall normalization and the power $\gamma_{\mathrm{sat}}$ is the (BFKL) saddle point in the vicinity of the saturation line $Q^2= Q_{\mathrm{sat}}^2(x)$. In addition, the  anomalous dimension is defined as $\gamma = 1- \gamma_{\mathrm{sat}}$. As usual in the BFKL formalism, $\bar{\alpha}_s=N_c\,\alpha_s/\pi$, $\beta \simeq 28\,\zeta (3)$ and  $Y=\ln (1/x)$. The quadratic diffusion factor in the exponential gives rise to the scaling violations.

The dipole cross section  in Eq. (\ref{sigmabfkl}) does not include an extrapolation from the geometric scaling region to the saturation region. This  has been recently implemented in Ref. \cite{iancu_munier}, where the dipole amplitude  ${\mathcal N} (x,\rr)=\sigma_{dip}/2\pi R_p^2$ was constructed to smoothly interpole between the  limiting behaviors analytically under control: the solution of the BFKL equation
for small dipole sizes, $\rr\ll 1/Q_{\mathrm{sat}}(x)$, and the Levin-Tuchin law \cite{Levin}
for larger ones, $\rr\gg 1/Q_{\mathrm{sat}}(x)$. A fit to the structure function $F_2(x,Q^2)$ was performed in the kinematical range of interest, showing that it is  not very sensitive to the details of the interpolation (For a comprehensive phenomenological analysis of the HERA results using the numerical solution of the BK equation see Ref. \cite{lubli}). The dipole cross section was parametrized as follows,
\begin{eqnarray}
\sigma_{dip}(x,\rr) = \sigma_0\, \left\{ \begin{array}{ll} 
{\mathcal N}_0\, \left(\frac{\rr\, Q_{\mathrm{sat}}}{2}\right)^{2\left(\gamma_{\mathrm{sat}} + \frac{\ln (2/\rr Q_{\mathrm{sat}})}{\kappa \,\lambda \,Y}\right)}\,, & \mbox{for $\rr Q_{\mathrm{sat}}(x) \le 2$}\,,\\
 1 - \exp^{-a\,\ln^2\,(b\,\rr\, Q_{\mathrm{sat}})}\,,  & \mbox{for $\rr Q_{\mathrm{sat}}(x)  > 2$}\,, 
\end{array} \right.
\label{CGCfit}
\end{eqnarray}
where the expression for $\rr Q_{\mathrm{sat}}(x)  > 2$  (saturation region)   has the correct functional
form, as obtained either by solving the Balitsky-Kovchegov (BK) equation \cite{bal,kov}, 
or from the theory of the Color Glass Condensate (CGC) \cite{jamal_iancu}. Hereafter, we label the model above by CGC. The coefficients $a$ and $b$ are determined from the continuity conditions of the dipole cross section  at $\rr Q_{\mathrm{sat}}(x)=2$. The coefficients $\gamma_{\mathrm{sat}}= 0.63$ and $\kappa= 9.9$  are fixed from their LO BFKL values. In our further calculations it will be used the parameters $R_p=0.641$ fm, $\lambda=0.253$, $x_0=0.267\times 10^{-4}$ and ${\mathcal N}_0=0.7$, which give the best fit result. A large $x$ threshold factor $(1-x)^5$ will be also considered, for sake of completeness.

Recently, this model has also been used in phenomenological studies of the vector meson production \cite{Forshaw1} and the diffractive processes \cite{Forshaw2} at HERA as well as hadron production in nuclear collisions at RHIC \cite{jamal}. Here we compare for the first time this model with the recent H1 data for the
longitudinal structure function.

\section{Results and Discussions}

\begin{figure}[t]
\psfig{file=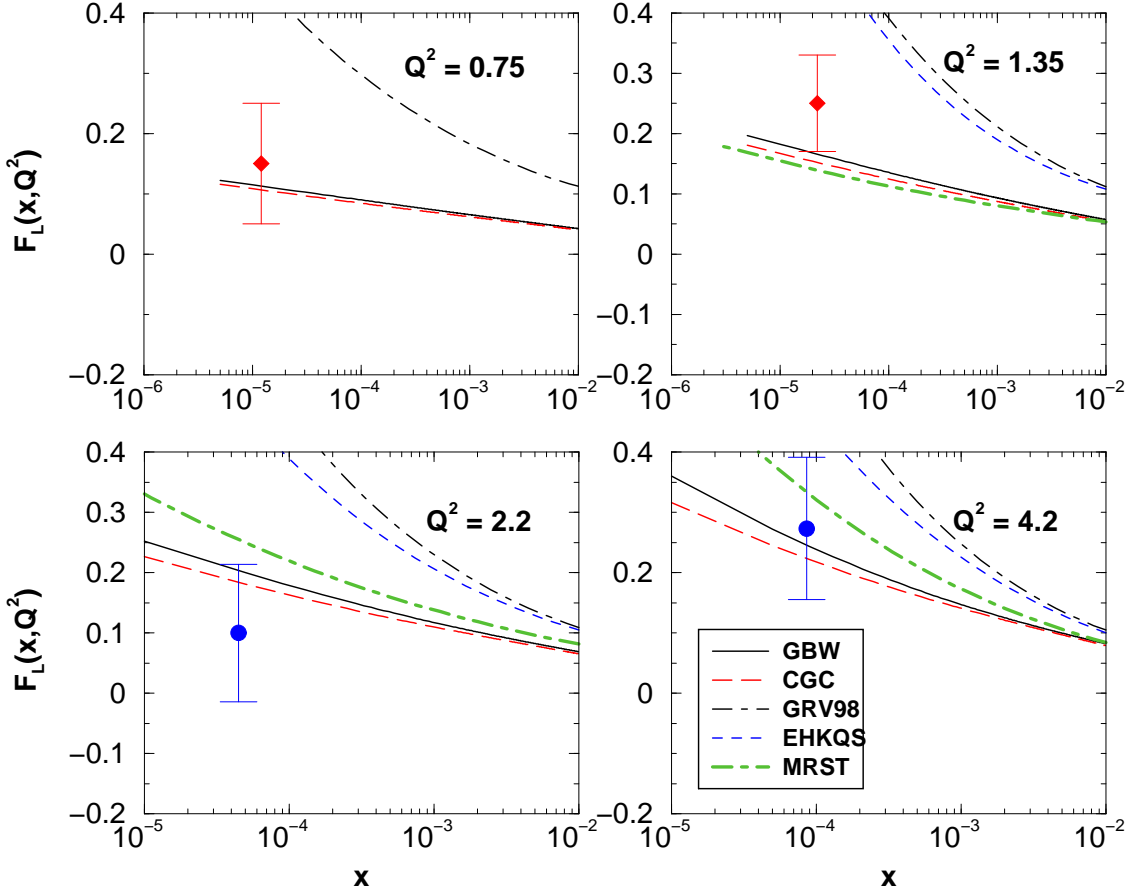,width=150mm}
 \caption{\it The results for $F_L(x,Q^2)$ as a function of $x$ at fixed low $Q^2$ values. The numerical results for the saturation models (GBW and CGC) as well as leading twist DGLAP approach for three inputs for the gluon distribution (GRV98, EHKQS and MRST) are presented. Data from H1 Collaboration. }
\label{fig1}
\end{figure}

\begin{figure}[t]
\psfig{file=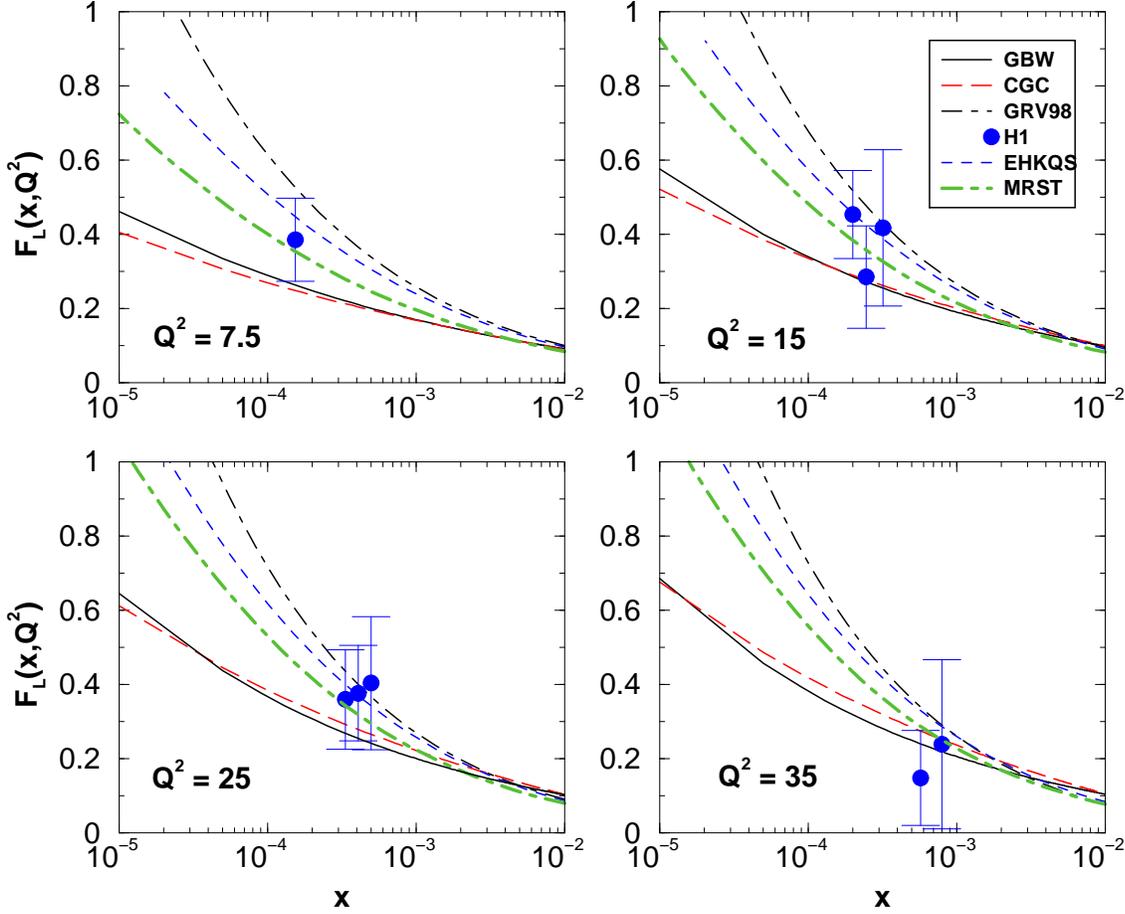,width=150mm}
 \caption{\it The results for $F_L(x,Q^2)$ as a function of $x$ at fixed large $Q^2$ values. Same notation of previous figure.}
\label{fig2}
\end{figure}

Lets present the numerical results coming out the saturation models (GBW and CGC) and from the usual collinear approach. In Figs. \ref{fig1} and \ref{fig2} the  predictions of various theoretical models
are compared with the H1 experimental results for the longitudinal structure function \cite{h1_prel,h1}. In particular, in Fig. \ref{fig1} we compare the distinct predictions with the recent H1 preliminary  $F_L$ data at $Q^2 = 0.75, \ 1.35$ GeV$^2$, as determined from the 1999
minimum bias and the 2000 shifted vertex H1 data. These points were extracted from the  plots on Ref. \cite{h1_prel}. The data show that $F_L$ remains non-zero down to the lowest $Q^2$ values
measured and already distinguish between the different models in
the low-$x$ region. The previous $F_L$ data \cite{h1} at larger $Q^2$ are also presented.

In Fig. \ref{fig1}, the lowest $Q^2$ bins are shown, in particular the new $Q^2=0.75$ and $Q^2=1.35$ GeV$^2$ measurements. At this region, the two saturation models (GBW and CGC) give very similar results as a consequence they have a similar behavior in the transition to the saturation regime. Namely, their main differences are in the large virtualities region, where CGC depends on the BFKL anomalous dimension at the saturation vicinity as referred before. The data description is very consistent, mostly at the lowest $Q^2$ points where the usual collinear approaches are unable to produce reliable results. For sake of comparison, a leading twist calculation is also presented. We have used Eq. (\ref{flalta}) and considered three different choices for the parton distributions. 
The predictions from the  GRV98   parameterization, which is obtained using the DGLAP evolution equation,  and the EHKQS gluon function \cite{Eskola}, which contains corrections from non-linear GLR  evolution equation, are not in agreement with experimental, even at large $Q^2$. The EHKQS gluon distribution slows down the dependence on $x$, but it is not enough to reach either to the upper limit of error bars. Notice that at $Q^2=0.75$ GeV$^2$ a DGLAP approach is unable to give reliable results, thought a backward QCD evolution is possible. For the GRV98 case, we have extrapolated the $Q^2=0.85$ GeV$^2$ initial condition down to $Q^2=0.75$ GeV$^2$ once they are very closer.  Similarly, for $Q^2 = 1.35$ GeV$^2$, we present the EHKQS prediction for $Q^2 = 1.4$ GeV$^2$ which is the lowest available $Q^2$ for this parameterization. On the other hand, the predictions for $F_L$ using the  MRST parameterization  reasonably describe the  H1 data, which is directly associated to the behavior  assumed for the  parton distributions in the initial evolution scale $Q^2 = 1.0$ GeV$^2$. In this case, the behavior of the sea distribution is independent of the gluon one, with the  input  gluon distribution being valence-like ($xg \propto x^{0.10}$), while  the input sea distribution  has a steep growth at small $x$ ($xS \propto x^{-0.19}$).

In Fig. \ref{fig2} one presents the high $Q^2$ analysis. Once again, the saturation models give a rasonable description of data, producing a milder $x$ growth than the DGLAP results. Their overall normalization also is smaller than the DGLAP analysis, becoming closer as virtuality increases. When comparing the  DGLAP results, one verifies that the intermediate $Q^2$ region at small-$x$ is an adequate kinematical region to study non-linear QCD corrections to the gluon distribution. There, very precise measuments of $F_L$ or large statistics could constraint the size of those corrections.

\begin{figure}[t]
\psfig{file=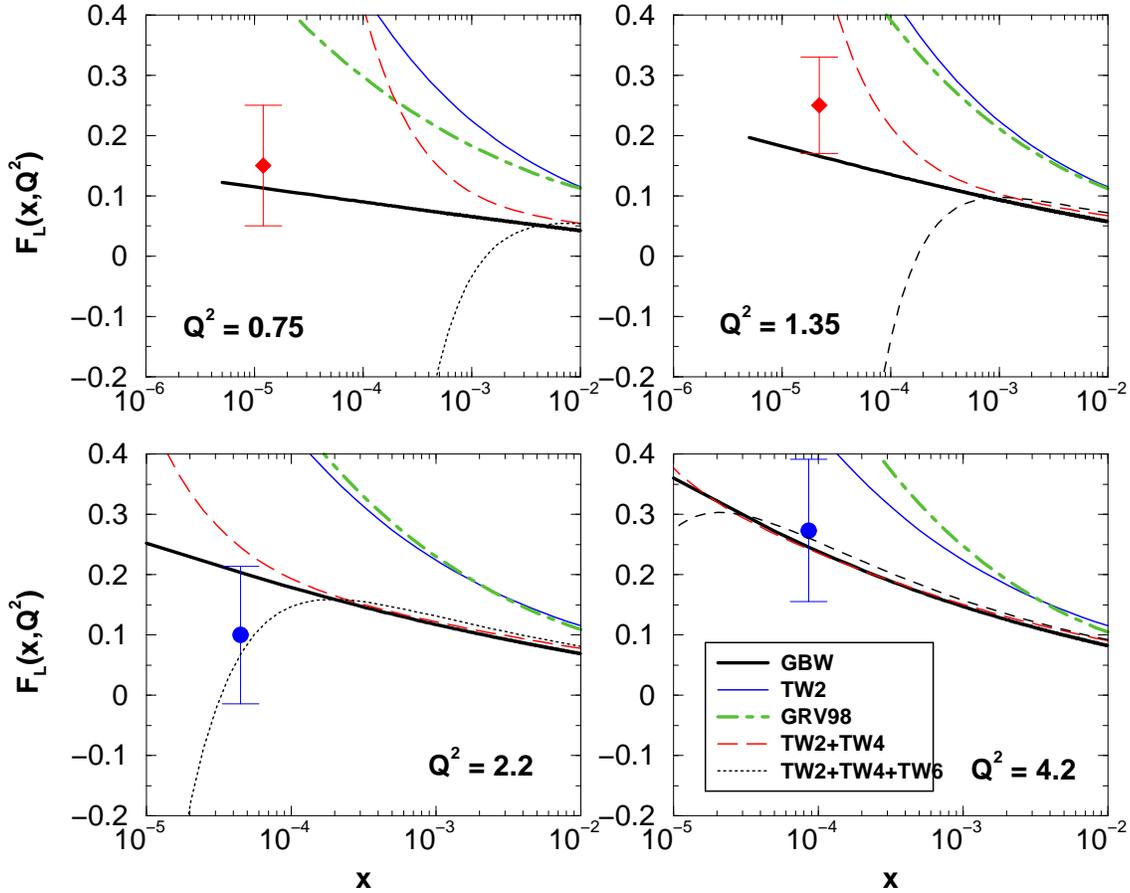,width=150mm}
 \caption{\it The results for $F_L(x,Q^2)$ as a function of $x$ at fixed large $Q^2$ values for the saturation model (GBW) and the sum of its twist contributions. The leading twist (twist-2) and the first 2 higher twist (twist-4 and twist-6) correction are presented. The leading twist DGLAP results (using GRV98 gluon pdf) is shown for comparison. }
\label{fig3}
\end{figure}

Finally, in Fig. \ref{fig3} we present a comparison among the
contributions of the different twists. In order to do this, we show separately 
the result from the summation of the different twists  for the analitically simple expression in Eqs. (\ref{twist4l}-\ref{twist6l}) 
for the saturation model.  
For comparison we also present the GBW prediction, which represents the full higher-twist ressumation. 
  In order to compare the present analysis with a leading 
order and leading twist calculation, we present in addition the results of Eq. (\ref{flalta})  considering the GRV98
gluon distribution. 
We can see  that the twist-2 contribution from the saturation model is in complete numerical 
agreement with the  DGLAP result, showing a consistent reproduction of leading twist contribution in its twist resummation. We have that the twist-4 and twist-6 terms gives important contributions, strongly modifying the magnitude the longitudinal structure function and its $x$-behavior. 
Moreover, the $x$ value  where the curves become distinct is $Q^2$-dependent, as expected from the energy dependence present in the saturation scale and consequently in the variable $\xi$. In particular, for $Q^2 = 4.2$ GeV$^2$ we have the sum of the twist-2 and twist-4 terms reasonable reproduces the GBW prediction for $x > 10^{-5}$, while for $Q^2 = 2.2$ GeV$^2$ this approximation is only valid for $x > 10^{-4}$. For smaller values of $Q^2$ only the full ressumation gives a good description of the experimental data, which demonstrate a twist summation term by term (i.e., summing the first contributing terms)  would be either incomplete in that kinematical regime.

As a summary, we have analyzed the longitudinal structure function at low $Q^2$ and small-$x$, which  directly depends on the gluon distribution function, within the saturation approach. In particular, we have show the saturation models (GBW and CGC) describe consistently the recent low $Q^2$ H1 data, even at very low virtualities. Moreover, we have presented the higher twist contributions, using the simple analytical expressions provided by the GBW model. It is verified that  they play an important role at small-$x$ for the virtualities considered here. Concerning the leading twist DGLAP analysis, which we have considered for sake of comparison, it is shown that non-linear GLR corrections to the gluon distribution function are not enough to bring the numerical analysis to the recent experimental results. However, we have shown the intermediate $Q^2$ region should be an important kinematical region where the size of such corrections could be investigated. In general lines, $F_L$ is an outstanding observable testing both parton saturation and twist resummation. Therefore, more precise data and/or more statistics is increasingly desirable at low $Q^2$ and small-$x$. Significant  further progress in $F_L$ measurements at HERA can
only be made by reducing the proton beam energy. A run with reduced proton beam energies is planned for the next few years.

\section*{Acknowledgments}
M.V.T.M. thanks the support of the High Energy Physics
Phenomenology Group, GFPAE IF-UFRGS, Brazil. This work was
partially financed by the Brazilian funding agencies CNPq and
FAPERGS.

\end{document}